\documentclass{ar-1col}
\usepackage{url}
\usepackage{setspace}
\usepackage[numbers]{natbib}

\usepackage{cancel}
\usepackage{amsmath}

\jname{Prepared for Annual Reviews of Biophysics}

\jyear{2024}
\doi{10.1146/annurev-biophys-081624-030543}

\begin{document}

\markboth{Russo et al.}{Soft modes across scales}


\title{Soft Modes as a Predictive Framework for Low Dimensional Biological Systems across Scales}


\author{Christopher Joel Russo,$^1$$^3$ Kabir Husain,$^1$$^4$ and Arvind Murugan$^1$$^2$\\[12pt]
\footnotesize\normalfont $^1$James Franck Institute, University of Chicago, Chicago, United States\\
$^2$Department of Physics, University of Chicago, Chicago, United States
\begin{spacing}{1} $^3$  Program in Biophysical Sciences, University of Chicago, Chicago, United States \end{spacing}
\begin{spacing}{1}$^4$Department of Physics, University College London, London, United Kingdom\end{spacing}}
\begin{abstract}
All biological systems are subject to perturbations: due to thermal fluctuations, external environments, or mutations. Yet, while biological systems are composed of thousands of interacting components, recent high-throughput experiments show that their response to perturbations is surprisingly  low-dimensional: confined to only a few stereotyped changes out of the many possible. Here, we explore a unifying dynamical systems framework - soft modes - to explain and analyze low-dimensionality in biology, from molecules to eco-systems. We argue that this one framework of soft modes makes non-trivial predictions that generalize classic ideas from developmental biology to disparate systems, namely: phenocopying, dual buffering, and global epistasis. While some of these predictions have been borne out in experiments, we discuss how soft modes allow for a surprisingly far-reaching and unifying framework in which to analyze data from protein biophysics to microbial ecology.
\end{abstract}

\maketitle

\tableofcontents

\section{INTRODUCTION}

Biological systems are, at face value, high dimensional. Proteins have dozens to thousands of amino acids, and may have hundreds of thousands of atoms. A single cell has thousands of genes, the expression levels of which all vary. Multicellular organisms or a microbial ecosystem can have trillions of cells, each with their own gene expression patterns. Yet, decades of empirical work has shown that the variation in the state of biological systems with many degrees of freedom can be described with far fewer dimensions than the number of degrees of freedom. 
\par The concept of dimensionality reduction in biology has been discussed widely in recent years \cite{eckmann2021dimensional, kaneko2024constructing}. However, the meaning and origin ascribed to this reduction has been diverse, ranging from biological or evolutionary constraints \cite{sato2020evolutionary}  to simply the  nature of high dimensional data \cite{moore2018high}. In this work, we will focus on one particular framework for understanding low dimensionality -- soft modes.
\par In the 1940s and 1950s, Waddington and Schmalhausen introduced concepts related to dimensionality reduction in the context of developmental biology: they argued that developmental processes push the morphological state of the developing organism down robust pathways (`canalization').  Through qualitative arguments (and famous drawings), they suggested several other non-trivial consequences\cite{schmalhausen1949factors, waddington1942canalization, waddington2014strategy} that have primarily been explored in a developmental context.

In this perspective, we wish to suggest that a mathematicization of these ideas, in terms of soft modes of a dynamical system, make these consequences relevant to biological organization across scales. In particular, we suggest three specific predictions for any level of biological organization with dimensionality reduction that can be captured by a soft modes framework:

\begin{enumerate}
    \item Phenocopying -  phenotypic variation due to by environmental changes will align with the variation due to mutational changes
    \item Dual buffering - molecular mechanisms evolved to buffer the impact of environmental stresses (e.g., heat shock pathways) will also buffer the impact of mutations 
    \item Global epistasis -  interactions between perturbations (e.g., mutations) at different sites, genes or elements at higher levels of organization will show constrained low dimensional patterns.
\end{enumerate}

We begin by highlighting empirical observations of low dimensionality that motivate a description in terms of soft modes. We then explain how the soft mode framework makes several non-trivial predictions, which have been empirically observed in many contexts. We conclude with a discussion of why soft modes may arise in biological systems across scales.

\section{Biological systems are often observed to be effectively low dimensional}

\subsection{Structural variation of proteins}

Many proteins have well defined 3-dimensional structures, which comprise the relative positions of their constituent atoms, that at least partly determines their function \cite{dunker2008function}. However, the 3-dimensional structure of a protein can vary, both across different environmental conditions (e.g., due to interactions with ligands or other molecules) and across a protein family (due to mutations). Experiments and theoretical modeling shows that structural variation in proteins due to environmental variation and mutations often effectively occupy a common low dimensional subspace of conformations (fig \ref{examples_fig}a).  

\par For example, Macias et al. \cite{leo2005analysis} quantified the dimensionality of structural variation within 35 protein families. They aligned experimentally resolved structures within each protein family, producing a matrix $\textbf{X}_{n \times 3p}$ of cartesian coordinates, where $n$ is the number of proteins in the family and $p$ is the number of residues. PCA on this matrix revealed that structural variation can be overwhelmingly be captured by 4 or 5 dimensions.

In parallel, they analyzed, \textit{in silico}, the modes of physical deformation of a single protein in each family. To do so, they employed elastic network models in which the $C_{\alpha}$ atoms are connected by harmonic potentials. A normal mode analysis of this model showed that some set of deformation modes had smaller eigenvalues $\lambda_i$, i.e., are softer than others -- changing the conformation along that mode is energetically easier than along other modes. The conformational dynamics of such protein, due to thermal fluctuations, ligand binding or other perturbations, will be dominated by dispalcement along these modes. 

Remarkably, these soft modes of a single protein, deforming as an elastic material due to forces, generally agree with the dominant deformation modes revealed by principal component analysis (PCA) on structures across a protein family due to mutations \cite{leo2005analysis}. Similar analyses comparing physical and evolutionary structural deformations have been performed in greater detail for particular protein families, such as globins \cite{Echave2010} and the Ras GTPase superfamily \cite{Raimondi2010}. We discuss the significance of this coincidence in the section on the `phenocopying' consequence of soft modes.  

\par With the growing number of structures in the PDB, as well as recent interest in protein dynamics, a number of other works have conducted similar analyses: quantifying the dimensionality of protein structural diversity either across a family, or for one protein in response to physical perturbations \cite{Sinha2002,Yang2008,Rod2003}. For environmental perturbations,  Amadei et al. \cite{amadei1993essential}  find a similar result using molecular dynamics simulation -- that most of the structural variation explored by lysozyme in solution is constrained to a few dimensions . Recently, more direct experimental probes of the deformation of proteins under physical forces have come online. Nuclear Magnetic Resonance (NMR) methods have been used to empirically investigate collective modes of deformation in proteins, and have identified low dimensionality in dynamic variation \cite{boehr2006dynamic}. The ability of NMR to capture structural variation across timescales is limited, however. Recent advances in time-resolved X-ray crystollgraphy (EFX) promise a direct empirical observation of the mechanical deformation modes of a protein on fast and slow timescales \cite{henning2024biocars} are providing direct experimental data on soft modes in protein dynamics \cite{hekstra2016electric}. Low dimensionality in the motion of proteins and its centrality in protein evolution and relation to protein function and allostery have been highlighted in other works \cite{brooks1983harmonic, eckmann2019colloquium, bahar2010functional}.

\subsection{Microbial gene expression patterns across environments} 
\begin{figure}[h]
\includegraphics[width=0.7\linewidth]{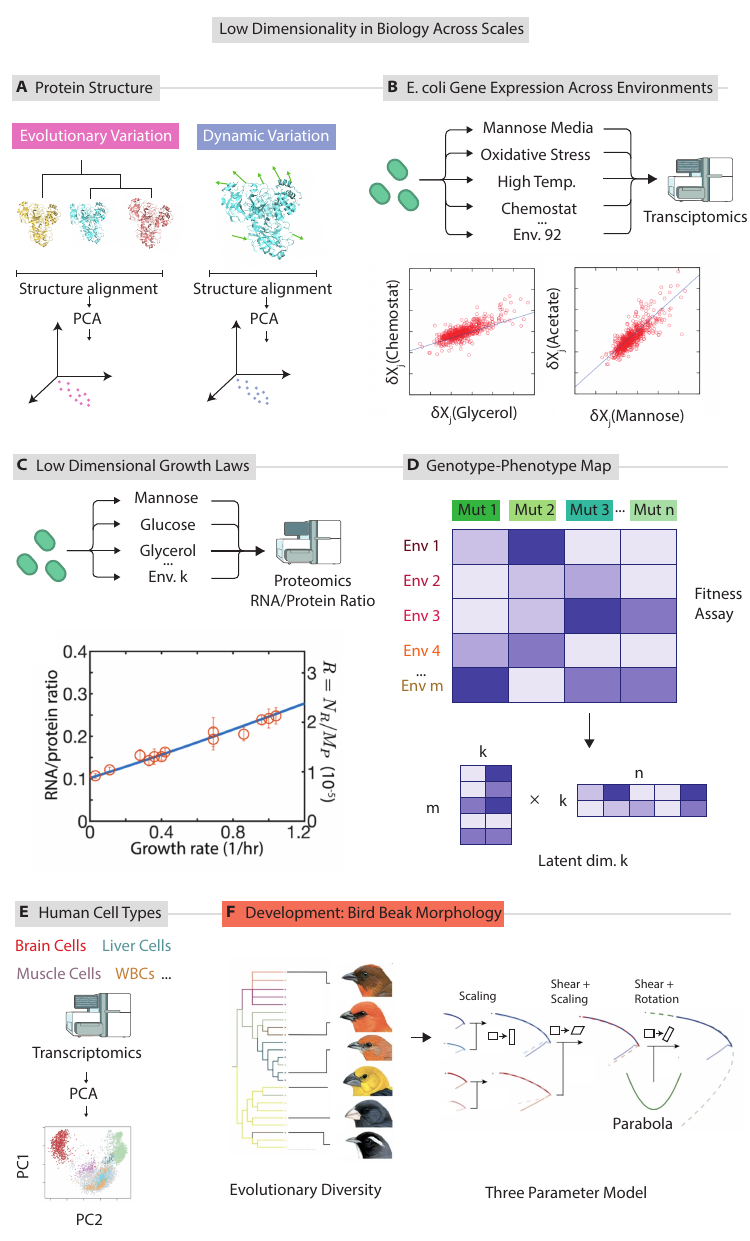}
\caption{\textbf{Examples of low dimensional variation in biological systems across scales.} (a) The dynamic variation of a single protein's structure and the evolutionary variation in structures across a protein family is low dimensional \cite{amadei1993essential,leo2005analysis,tang2021dynamics}. (b) Variation of the transcriptomic state $\vec{\delta E}_i$ of \emph{E. coli} grown in various environmental conditions $i$ lies in a low dimensional subspace compared to the number of transcripts \cite{furusawa2018formation,sastry2019escherichia,tsuru2024genetic,stewart2012cellular}. (c) Coarse-grained proteome allocation of \emph{E. coli} across high-dimensional variation of environment (e.g., nutrients, drugs) can be explained by one number -- growth rate.\cite{wu2022cellular,scott2014emergence}. (d) The genotype-to-phenotype map for many mutants of yeast across many environmental conditions can often be factored through a low dimensional latent phenotype space (dimension $k=8$ example shown) \cite{kinsler2020fitness}. (e) The gene expression state of diverse kinds of human cell types lie on a low dimensional manifold, as revealed by PCA of transcriptomics data \cite{lukk2010global}. (f) Morphological variation in body parts across related species tends to be low dimensional compared to the naive dimensionality of phenotype space \cite{fritz2014shared}.}
\label{examples_fig}
\end{figure}

The gene expression state of a bacterial cell is very high dimensional, with thousands of genes under regulation, with their expression varying over time in response to changes in the environment. Mass spectroscopy\cite{cox2011quantitative} and RNA sequencing\cite{lowe2017transcriptomics} allows for the high throughput quantification of gene expression across many genes (Fig. \ref{examples_fig}b).

Empirically, it has been observed that although thousands of genes have varying expression, the variation in expression state --- as seen through (bulk) RNA sequencing and proteomics --- across many environments and across mutants can often be captured by a small number of degrees of freedom \cite{furusawa2018formation,sastry2019escherichia,tsuru2024genetic,stewart2012cellular,matsumoto2013growth,schmidt2016quantitative}. 
In one approach\cite{furusawa2018formation,kaneko2015universal}, 
the vector $\delta \vec{x}(E_i)$ -- the change in gene expression state from the baseline, unperturbed condition to environmental condition $E_i$ -- is computed:
        $$\delta \vec{x}(E_i) = \vec{X}(E_i) - \vec{X}(E_0)$$

PCA on a matrix of $\delta \vec{x}(E_i)$ across many experiments $E_i$ reveals a few dominant modes that are fewer in number than the number of environments (results plotted in Fig. \ref{examples_fig}b).  Environmental conditions might include osmotic stress, starvation, and heat stress, at varying levels, varying carbon source in batch growth, and grown in chemostat and batch at various growth rates. Similar analyses are done for mutational changes, typically using mutants that arise in long term evolution experiments \cite{tsuru2024genetic, maeda2020high,stewart2012cellular,sastry2019escherichia}.

\subsection{Phenotypic variation among cell types}
 Multicellular organisms can have diverse cell types, differentiated from common ancestral stem cells. Many studies that quantify gene expression across these lineages summarize the expression differences with a small number of variables, employing dimensionality reduction techniques such as t-SNE, PCA, or UMAP (Fig. \ref{examples_fig}e)\cite{lukk2010global,neu2017single,xie2021single,kobak2019art,kulkarni2019beyond}. Work since the early 2010s has identified this low dimensionality in mammalian cell types \cite{lukk2010global}, and many other papers since have found similar results \cite{gadaleta2011global,zhou2018data,schneckener2011quantifying}

\par 
This fact of the low dimensionality of gene expression across cell types predicts that relatively few inputs are likely necessary to reprogram cell types -- if the diversity of cell types lies on $k$ dimensional manifold, turning $k$ knobs may be sufficient to move the cell along the manifold \cite{muller2011few}. This prediction has been born out in empirical work, with 1 to 5 perturbations necessary to change mammalian cell type \cite{huang2011induction} \cite{szabo2010direct}. Smart et al. \cite{smart2023emergent} have argued that the low dimensionality of cell states is an emergent property of regulatory networks with cell to cell couplings, showing that simple Hopfield models of intracellular regulation with couplings between cells leads to few attractor states. 
\par The use of these dimensionality reduction methodologies in these contexts have come under criticism, with some arguing that the flexibility of these methods leads spurious dimensionality reduction\cite{chari2023specious}. However, it is still likely true that the effective dimensionality of differentiated cells is much lower than the number of genes under regulation.

\subsection{Bacterial Growth Laws}
Bacteria regulate the expression of different proteins based on the conditions they grow in; for example, conditions lacking a nutrient might require higher expression of a specific metabolic enzyme. While genes are regulated in such environment-specific manners, work over several decades has revealed a layer of global regulation that is simpler \cite{schaechter1958dependency, maaloe1979regulation,klumpp2009growth}. 

When the proteome is divided into sectors corresponding to ribosomal proteins, metabolic and uptake proteins and other housekeeping proteins, the overall expression level of these sectors (or finer grained sectors, as shown in \cite{schmidt2016quantitative}) does not depend on all the high dimensional details of the environmental conditions \cite{schaechter1958dependency,klumpp2009growth} but instead depends on the high dimensional environment through a low-dimensional representation. When nutrients are varied, the growth rate achieved by the bacteria in that environment exactly predicts sector allocation. (Fig. \ref{examples_fig}c) \cite{schaechter1958dependency,klumpp2009growth,wu2022cellular,scott2014emergence,maaloe1979regulation}. 

The mechanistic origin of such low dimensional control lies in systems like ppGpp feedback\cite{wu2022cellular} and cAMP \cite{you2013coordination} that resemble a controller that, in effect, listens to a single (or a few) state variable(s) that integrates information about the metabolic state of the cell and thus regulates response to diverse environmental changes. If these simple controllers that allocate resource between sectors are also the dominant mechanism controlling within-sector expression levels, these growth laws will lead cells' gene expression states to lie on a low dimensional subspace.  The dimensionality of this subspace will determined by the number of simple controllers.

\subsection{Genotype-Phenotype Maps}
Genotype-phenotype maps built from high-throughout data have often found to factor through a lower dimensional latent space  (i.e., possess a low dimensional structure) \cite{petti2023inferring,sato2023prediction, tlusty2017physical,suzuki2014prediction,kinsler2020fitness}, even though both genotype and phenotype are high dimensional. For example, Kinsler et al \cite{kinsler2020fitness} assayed the fitness of $\sim 292$ yeast mutants in $\sim 45$ diverse environments. 

The authors then attempted to build a model of genotype to phenotype mapping where both the strain and the environment lie in a common $k$ dimensional phenotype space (Fig. \ref{examples_fig}d). That is, each mutant $i$  is assumed to be represented by a $k$ dimensional phenotype vector $(p_{i1},p_{i2},...p_{ik})$ while each environment $j$ is represented by vector $(e_{j1},e_{j2},...e_{jk})$ in the same $k$ dimensional latent space. The fitness of strain $\vec{p_i}$ in environment $\vec{e_j}$ is assumed to be $\vec{p_i}\cdot\vec{e_j}$. Using SVD, they find that a model with $k = 8$ is sufficient to predict the fitness of 292 mutants across the 45 diverse environments. 
This analysis found that mutations were typically pleitropic, having components across several of the $k$ dimensions. And yet, there are just $k \sim 8$ traits or features of the genotype and phenotype, each of which are naively much higher dimensional, that explain fitness.

\par Similar results have been obtained in several other contexts \cite{tlusty2017physical,petti2023inferring,suzuki2014prediction,sato2023prediction}. For antibiotic resistance, in \cite{suzuki2014prediction}  investigated a number of \textit{E. coli}  mutants' fitness exposed to diverse antibiotic perturbations. They define cross resistance $\mu(\vec{e_1},\vec{e_2}) $ as the fitness of a mutant that has evolved in environment $\vec{e_2}$ in environment $\vec{e_1}$. They show that $\mu(\vec{e_1},\vec{e_2}) $ can be approximated by $\tilde \mu(\delta y_1,...\delta y_D)$ where $\delta y_i$ is $i$th principal component of the difference between the gene expression state of a mutant adapted to $\vec{e_1}$ and a mutant adapted to $\vec{e_2}$. They find $D=3$ principal components are necessary, showing antibiotic resistance lies on a low-dimensional phenotype space.
\cite{sato2023prediction}. 
These and other works, which find low-dimensional genotype-phenotype maps in protein structure \cite{tlusty2017physical} and human cell line genotoxin robustness \cite{petti2023inferring} provide more evidence that evolution may be happening in a lower dimensional space than the apparent complexity of living systems suggests.

\subsection{Low dimensional patterns in genetic epistasis}

Epistasis is the dependence of the fitness effect of a mutation on its context, e.g., the fitness effect of mutation at residue or gene locus $i$  may be different in the context of mutation at residue or gene locus $j$. The epistatic coupling between two such residues or genomic loci is defined as $\Delta\Delta F_{ij} = \Delta F_{ij} -  \Delta F_{i} - \Delta F_{j}$ where $\Delta F_{i}$ and $\Delta F_{j}$ are the changes in fitness (or any other trait $F$) due to a mutations at site $i$ and $j$ respectively, done in isolation; $\Delta F_{ij}$ is the change in fitness upon making mutations at both site $i$ and $j$ (Fig. \ref{epistasis_fig}a)\cite{starr2016epistasis,domingo2019causes}. If epistasis $\Delta\Delta F_{ij}$ is non-zero, then $\Delta F_{ij} \neq \Delta F_{i} + \Delta F_{j}$; that is, the impact of mutations at sites $i,j$ is not additive. Epistasis is accordingly also often described as a non-linearity in the genotype-to-function map. In general, one should expect epistatic coefficients between different pairs of sites $i,j,k,\ldots$ to be relatively unconstrained and independent.

Nevertheless, surprisingly simple mathematical relationships have been shown to accurately predict the context dependence of mutations.\cite{olson2014comprehensive,sarkisyan2016local,chen2022environmental,kryazhimskiy2014global,wang2019revealing}. Such constrained low dimensional patterns of epistasis have been labeled `non-specific epistasis' \cite{park2023simplicity,chen2022environmental} or  `global epistasis' \cite{husain2020physical,phillips2021binding,kryazhimskiy2014global,otwinowski2018inferring}  across these fields, though their precise definition and potential origin are likely distinct. 
 In general, a number of works have argued that epistatic couplings between different sites, in proteins and in other systems, are mediated by the impact of the mutations on a few underlying latent variables \cite{olson2014comprehensive,sarkisyan2016local,otwinowski2018inferring, park2023simplicity,gong2013stability,chen2022environmental,johnson2023epistasis,diaz2023global}.

Similar low dimensional patterns have been seen at the genomic scale \cite{kryazhimskiy2014global, johnson2019higher,diaz2023global,wei2019patterns} 
. In \cite{kryazhimskiy2014global}, the fitness effect of a given mutation in yeast in different backgrounds was found to strongly predicted by just the  baseline fitness of the background genotype, regardless of the details of the specific genotype (Fig. \ref{epistasis_fig}e) Such low dimensional epistatic patterns are thought to make 
evolution more predictable \cite{johnson2023epistasis,kryazhimskiy2014global}. Similar results have been discussed recently for composition-function relationships in microbial ecosystems \cite{diaz2024global}. We discuss low dimensional epistasis further as a consequence of soft modes later in this review. 

\subsection{Microbial ecology}

One basic perspective on low dimensionality in ecology is revealed by the relationship of Lotka-Volterra models and Consumer Resource models\cite{cui2024houches, chesson1990macarthur}. In the former, a community of $N$ species is parameterized by $O(N^2)$ interaction coefficients $A_{ij}$, i.e., each species $i$ can interact with each of other species $j$ through an independently chosen interaction $A_{ij}$. In contrast, in a Consumer Resource model, interactions between $N$ species are mediated through the interactions of those $N$ species with $K$ resources (e.g., nutrients).  Consequently, interactions between different pairs of species are not independent. In fact, integrating out the resources in a Consumer Resource model leads to a Lotka-Volterra model but whose interaction matrix $A_{ij}$ is of rank $K$, i.e., is effectively lower dimensional if $K < N$.

A distinct notion of low dimensionality concerns the map between species composition and ecosystem function for microbial communities, an ecological analog of the genotype-phenotype map for individual organisms \cite{skwara2023statistically, sanchez2023community}. Here, ecosystem function typically refers to an activity like net butyrate production from a feedstock or pathogen invasion resistance that is thought to require cooperative action by a whole community. These maps, inferred from high throughput data on many combinatorially assemled communities, also appear to factor through low dimensional latent spaces \cite{skwara2023statistically}. Analogs of epistasis, i.e., measuring the impact of species alone or in combination, show evidence of low dimensionality (global epistasis) \cite{diaz2024global}, paralleling epistasis at the genomic and protein levels \cite{ diaz2023global}.

Finally, the variation of species composition in ecosystems across different environments reveals signatures of low dimensionality \cite{shahin2023embed, plata2023phenomenological, krishnamurthy2022smbiot,plata2024designing,human2012structure}; species abundances do not vary independently but instead, collections of species show coordinated modes of covariation that form a lower dimensional space in species abundance space (often termed eco-modes, ecological normal modes or other equivalent terms \cite{shahin2023embed,frentz2015strongly, hekstra2012contingency}).

\subsection{Neural activity}

Neural firing patterns have often been observed to be lower dimensional than naive expectations \cite{feng2023activity}. The firing rate patterns of $N$ neurons should naively occupy generic parts of $2^N$ state space but in many contexts, observed states are much lower dimensional\cite{izhikevich2007dynamical} such as measurements from hippocampus, with firing from place cells\cite{hopfield2010neurodynamics, yoon2013specific, ocko2018emergent}. These are often interpreted in terms of low dimensional `continuous attractors' encoded by the connectivity patterns between the neurons and reflect the low dimensional environment.  The attractor dynamics in these neural systems reduce the effective dimensionality in much the same way as soft modes in  physiological systems do. We do not go into dimensionality reduction in neural systems further since they have been discussed before; further, the relevance of the key consequences of low dimensionality we focus on (phenocopying, dual buffering and global epistasis) is not clear. 

\section{Soft modes as a quantitative framework for  dimensionality reduction}
Here, we introduce the idea of soft modes in dynamical systems. In later sections we argue that this elementary idea from the theory of dynamical systems makes sharp and quantitative predictions about low dimensional biological systems of the type reviewed above. We begin with a dynamical system defined by, 
\begin{equation}
    \frac{d \vec{x}}{dt} = \vec{f}(\vec{x})
\end{equation}
where $\vec{f}$ is a function of the state of the system $\vec{x}$ in $n$ dimensions. 
Taylor expanding $\vec{f}(\vec{x})$ around a fixed point $\vec{x}_0$,

$$\frac{d \vec{x} }{dt} = \vec{f}(\vec{x}_0 + \delta   \vec{x} (t))\approx \cancelto{0}{\vec{f}(\vec{x}_0)} + \Big[ \frac{\partial \vec{f} }{\partial x_1} ...\frac{\partial \vec{f} }{\partial x_n}\Big]\delta \vec{x}(t)$$

where $\Big[ \frac{\partial \vec{f} }{\partial x_1}...\frac{\partial \vec{f} }{\partial x_n}\Big]$ is the Jacobian matrix $\textbf{J}_x$ of $\vec{f}(\vec{x})$ with respect to $x$. We then diagonalize $\textbf{J}_x$, finding the set of eigenvalues $\lambda_1,...\lambda_n$ and their corresponding eigenvectors $\vec{v}_1,...\vec{v}_n$. 
Rewriting $\delta \vec{x}(t)$ as a sum of its projections onto the eigenvectors of $\textbf{J}$
$\delta\vec{ x} = \sum c_i(t) \vec{v}_i.$ and solving for the dynamics,

we see that 
$$\vec{x}(t) = \vec{x}_0 + \sum e^{\lambda_i t} c_i(0)\vec{v}_i.$$

\begin{figure}[h]
\includegraphics[width=5in]{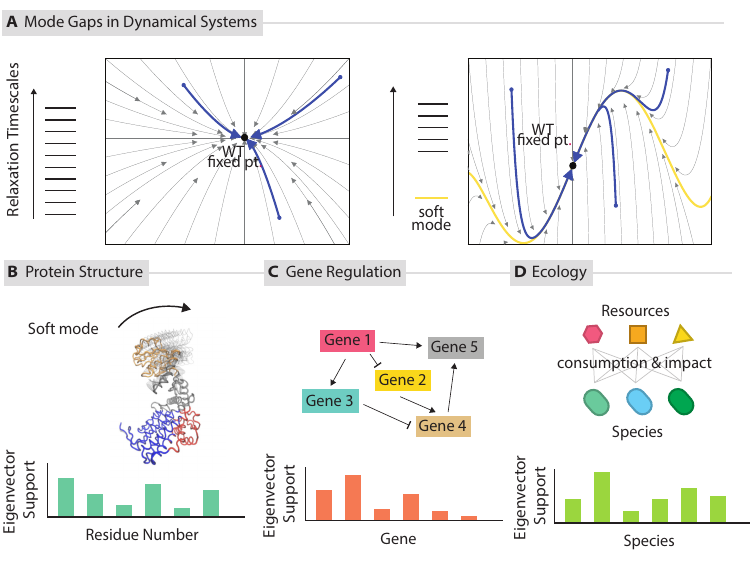}
\caption{ 
\textbf{Soft modes in dynamical systems.} (a) Perturbations away from a fixed point in a dynamical system decay with a spectrum of timescales..  Shown is a generic spectrum of relaxation timescales (left) vs a spectrum with a `mode gap' (right), i.e., a significant separation of timescales with one (or a few) slow relaxation modes. With a `mode gap', distinct perturbations in high dimensional state space decay quickly to the manifold associated with the soft modes (red), and slowly relax along that slow manifold to the fixed point. 
(b)  Soft modes in a protein (as revealed through Elastic Network Modeling (ENM), MD simulations, or other methods) represent `soft' mechanical deformations that are energetically less expensive than other deformations. (c) soft modes in a gene regulatory network correspond to those correlated patterns of gene expression that take the longest time to relax back to homeostasis. (d) Ecosystems with many distinct interacting species (e.g., microbial ecosystems) may also have slow dynamical modes in the high dimensional space of relative species abundance. }
\label{mode_gap_fig}
\end{figure}

If the fixed point is stable, all eigenvalues will have negative real parts. If some values $\lambda_i$ are dramatically smaller than other, then those modes will come to dominate the dynamics of the system on longer timescales, as the other modes quickly decay away (Fig. \ref{mode_gap_fig}a). Let us assume that $\lambda_i$ are sorted in descending order of $\operatorname{Re}  \lambda_i$. Then, for example, if $|\operatorname{Re}   \lambda_1 |<< |\operatorname{Re} \lambda_2|$, then at long timescales $t$ between $1/|\operatorname{Re} \lambda_1|$ and $1/ |\operatorname{Re} \lambda_2|$,
$$\vec{x}(t) \approx \vec{x}_0 +  e^{\lambda_1 t} c_1(0)\vec{v}_1$$

In general, modes with the smallest $\lambda_i$ and are the last to decay, are called \textbf{slow modes} or \textbf{soft modes} depending on the context.  If there is a large gap between $1/\lambda_1$ and $1/\lambda_2$, then the dynamics will very quickly become one dimensional, or $k$ dimensional if there are $k$ soft modes, i.e., $|\operatorname{Re}   \lambda_k |\ll |\operatorname{Re}   
 \lambda_{k+1}|$.

The spectrum for some dynamical systems might have a clear separation between fast and soft modes; i.e., $|\operatorname{Re}   \lambda_k |\ll |\operatorname{Re}   
 \lambda_{k+1}|$ with all $\lambda_i$ with $i\leq k$ being of a similar magnitude.   The ratio $|\operatorname{Re}   
 \lambda_{k+1}|/|\operatorname{Re}   
 \lambda_{k}|$ of timescales separating the (many) fast and (relatively few) soft modes is called the \textbf{mode gap}. While we assume a clear mode gap for ease of discussion here, many real systems may not have such a clear separation of timescales.  The three consequences presented in later sections here will likely qualitatively apply to spectra that spaced out, e.g., sloppy spectra \cite{brown2003statistical,gutenkunst2007universally}
, but the theoretical work and comparisons to data have not been done. 

\par
In a biological system, the fixed point $\vec{x}_0$ could represent a physiological optimum in gene expression state space for example, at which homeostatic mechanisms try to keep expression levels or metabolite levels fixed. However, this framework also applies to the dynamics of systems with no fixed points but rather stable trajectories that are attractors such as limit cycles. The modes of decay as the state trajectory approaches the limit cycle can also display a similar separation of timescales\cite{chachra2012structural, izhikevich2007dynamical}.

\par  
This framework also applies to non-equilibrium statistical ensembles described by a Master Equation \cite{husain2020physical}\cite{Qian2006-ag} , $\partial_t p_i = \sum_j k_{ij} p_j - \sum_j k_{ji} p_i \equiv \sum_j W_{ij} p_j$. Here, $p_i$ is the probability of a stochastic system occupying state $i$ while $k_{ij},k_{ji}$ are rates of transitions between different states. These models can be used to describe fluctuating molecular ensembles, including those that break detailed balance (e.g., a molecular motor or active enzyme). In such continuous-time master equation systems, matrix $W$ always has a zero eigenvalue corresponding to the steady state probability distribution and not to a decay timescale; hence, the relevant mode gap to judge whether the system has a soft mode in question refers to the ratio of timescales of the second and third smallest eigenvalues.

Finally, an important special case of this framework are systems whose dynamics $\vec{f}(\vec{x}) = - \vec{\nabla} V(\vec{x})$ follow from a well-defined energy function $V(x)$. For example, the deformations of most protein structures can be modeled in this way. In this case, the prior description of eigenvalues $\lambda_i$ reduces to a normal mode analysis of the energy function $V(\vec{x})$, expanded near a stable point $\vec{x}_0$,
\begin{equation}
    V(\delta \vec{x}) =   \frac{1}{2}\sum_i \sum_j \frac{\partial^2 V}{\partial x_i \partial x_j} \delta x_i \delta x_j
\end{equation}
where $\delta \vec{x} = \vec{x} - \vec{x}_0 $. The prior eigenvalues $\lambda_i$ are given by the eigenvalues of the Hessian matrix $\textbf{H} = \frac{\partial^2 V}{\partial x_i \partial x_j}$. Thus in energy-based systems, especially mechanical ones like proteins, soft modes can be intuitively understood as the least energetically expensive deformation modes.

\section{Consequences and Predictions}
\subsection{Phenocopying}
\begin{figure}[h]
\includegraphics[width=0.68\linewidth]{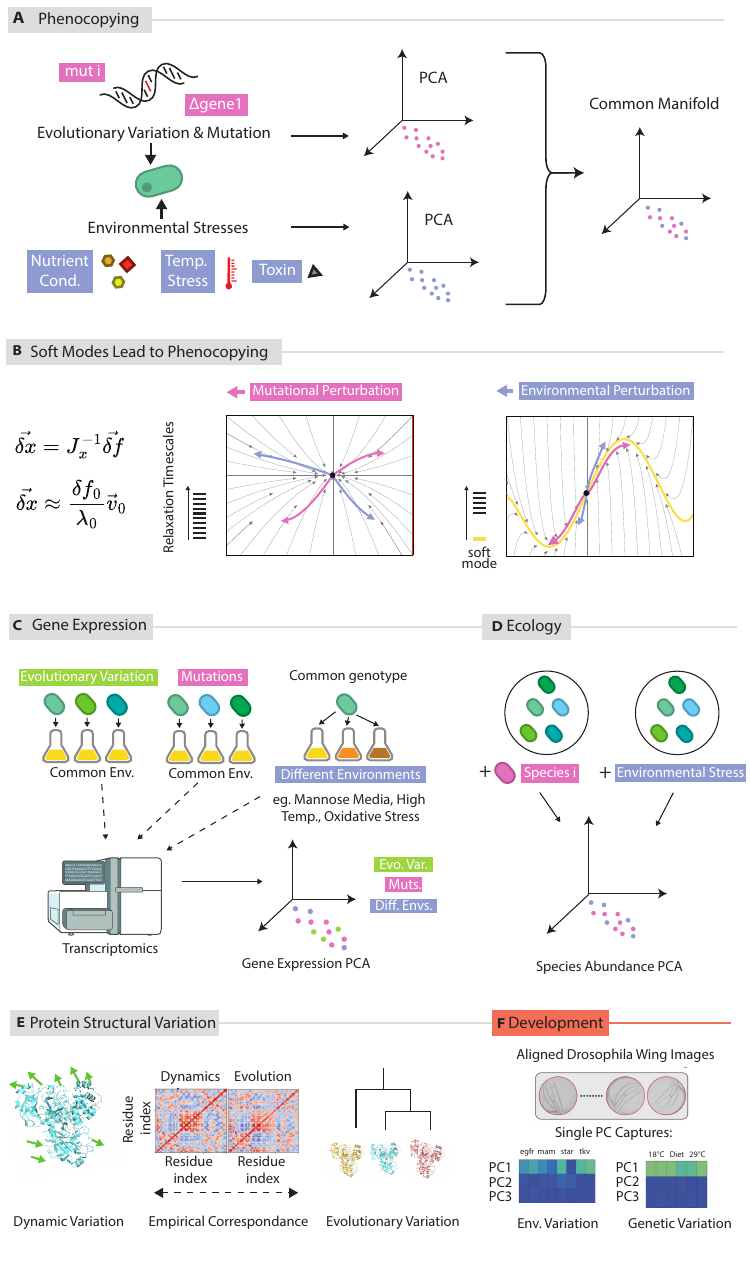}
\caption{\textbf{Soft modes predict phenocopying.} (a) Environmental perturbations are said to \textit{phenocopy} genetic mutants when they both lead to similar phenotypic outcomes, often lying along a common low dimensional manifold. (b) A system with a significant mode gap will show phenocopying as both environmental or mutational changes will dominantly perturb the system along the soft mode. See Eqn. \ref{phenocopy1} and \ref{phenocopy2}.  (c)  Phenocopying likely manifests in ecological systems, which are thought to exhibit soft modes. Here, the addition of a new species to a community is analogous to a mutation. (d) Gene expression state due to both environmental stresses and mutations in \textit{E. coli} map to a common low dimensional space. (e) Evolutionary variation in protein structure (i.e., due to mutations across a protein family) and dynamic variation (i.e., due to forces causing physical deformations in one protein) lie along one common low dimensional manifold\cite{amadei1993essential}. Plot from \cite{tang2021dynamics}.(f) Phenocopying was first described in developmental contexts (see \cite{raju2023theoretical} for a modern perspective). In recent work by Mani et al \cite{alba2021global}, alignment of wing images from genetically diverse as well as those that developed under external stresses reveals a common low dimensional space of variation. }
\label{phenocopy_fig}
\end{figure}

Phenocopying refers to a duality between phenotypic variation caused by mutations and environmental perturbations (Fig. \ref{phenocopy_fig}a), first described in developmental biology contexts \cite{ goldschmidt1935gen,mitchell1978heat}.
In development, phenocopying suggests that morphological variation in embryos due to environmental perturbations during development (e.g., exposure to toxins, temperature changes) will reflect morphological variation due to mutational perturbations to the developmental process (Fig. \ref{phenocopy_fig}f) \cite{goldschmidt1935gen, sawin1932hereditary,alba2021global,mitchell1978heat, raju2023theoretical}.

\par Here, we argue that phenocopying is a natural and immediate consequence of soft modes in a dynamical system; as a consequence (Fig. \ref{phenocopy_fig}b), phenocopying is relevant to contexts beyond development with striking experimental predictions at scales ranging from proteins to microbial ecology. 
We begin again with a dynamical system defined by
$
    \frac{d \vec{x}}{dt} = \vec{f}(\vec{x},\vec{e}, \vec{g})
$ where the parameter vectors $\vec{e}$ and $\vec{g}$ reflect the environmental state and genome respectively. 
As earlier, we assume that the system has a fixed point $\vec{x_0}$ for ``wildtype'' parameters $\vec{e_0}$ and $\vec{g_0}$, $\frac{d \vec{x}}{dt} = f(\vec{x}_0,\vec{e_0}, \vec{g_0}) = 0$. 

In response to an environmental perturbation $\vec{e_0} + \delta{\vec{e}}$, the fixed point $\vec{x}_0$ will be perturbed by $\delta_e \vec{x}$; in a linearized approximation, $\delta_x \vec{x}$ is given by 
\begin{equation}
      \delta_e \vec{x} =\mathbf{J}_x^{-1} \textbf{J}_e \delta\vec{e} =  \sum_i \frac{\big(\textbf{J}_e \delta \vec{ e}\big) \cdot \vec{v_i}}{\lambda_i} \vec{v}_i
\end{equation}
where $\textbf{J}_e =  \Big[ \frac{\partial \vec{f} }{\partial e_1}...\frac{\partial \vec{f} }{\partial e_n}\Big]$ and as above, $\lambda_1,...\lambda_n$ are the eigenvalues and the corresponding eigenvectors $\vec{v}_1,...\vec{v}_n$ for $\textbf{J}_x$.

If we follow the same logic for perturbations in state $\delta_g \vec{x}$ due to a mutational perturbation $\vec{g}_0 + \delta{\vec{g}}$, we find,
\begin{equation}
      \delta_g \vec{x} =\textbf{J}_x^{-1} \textbf{J}_g \delta\vec{g} =  \sum_i \frac{\textbf{J}_g \delta\vec{g} \cdot \vec{v_i}}{\lambda_i} \vec{v}_i
\end{equation}
where now $\textbf{J}_g =  \Big[ \frac{\partial \vec{f} }{\partial g_1}...\frac{\partial \vec{f} }{\partial g_n}\Big]$.

In general, the ensemble of variation $\{ \delta_{e1} \vec{x},\delta_{e2} \vec{x}... \}$ created by many different environmental perturbations $\{ \delta \vec{e}_1,\delta \vec{e}_2...\}$ is unrelated to the ensemble of variation $\{ \delta_{g1} \vec{x},\delta_{g2} \vec{x}... \}$ created by many different mutational perturbations $\{ \delta \vec{g}_1,\delta \vec{g}_2...\}$; in particular, the two ensembles depend on $\textbf{J}_e$ and $\textbf{J}_g$ vectors respectively; these vectors encode how mutations and environmental factors microscopically affect the state of the system and are idiosyncratic and not related to each other in general.

However, in the presence of a soft mode, $\lambda_1 \ll \lambda_i$ for $i>1$. 

In this limit, the details of $\textbf{J}_e$ and $\textbf{J}_g$ do not matter as long as $\textbf{J}_e \delta\vec{ e}$ and $\textbf{J}_g \delta\vec{ g}$ have a non-zero projection in the direction of the soft mode $\vec{v_1}$. In this limit,  
\begin{equation}\label{phenocopy1}
      \delta_g \vec{x}  \approx \frac{\textbf{J}_g \delta\vec{g} \cdot \vec{v_1}}{\lambda_1} \vec{v}_1
\end{equation}
\begin{equation}\label{phenocopy2}
      \delta_e \vec{x} \approx   \frac{\textbf{J}_e \delta\vec{e} \cdot \vec{v_1}}{\lambda_1} \vec{v}_1
\end{equation}
That is, environmental and mutational perturbations will have the same stereotyped effect on the state of the system, along the vector $\vec{v_1}$.  In realistic cases where there are some small number of soft modes $n$, the ensemble $ \{ \delta_{gi} \vec{x} \} $ of structural variation seen due to environmental changes will occupy the same  $n$ -dimensional subspace as the ensemble $ \{ \delta_{ei} \vec{x} \} $ of structural variation due to mutational changes. Here, $n$ can still be much smaller than the naive dimensionality $N$ of $\vec{x}$ (e.g., number of residues in a protein, genes in a organism, microbial species in an ecosystem etc). 

\par \textbf{Consequences:}
Since phenocopying is intrinsic to soft modes, we predict phenocopying in contexts outside of developmental biology as already seen in some empirical work. 

 \par Protein structural variation displays low-dimensional phenocopying (Fig. \ref{phenocopy_fig}e) \cite{leo2005analysis, tang2021dynamics}. The low dimensional subspace that contains protein structural variation across a family --- due to mutations --- maps to precisely the same subspace that contains variation due to environmental perturbations. Structural variation seen in solution (due to physical fluctuations) using NMR \cite{best2006relation} is similar to structural variation across closely related homologs (due to mutations). In addition, AlphaFold and related structure prediction methodologies that exploit evolutionary data to predict the structure of unsolved sequences have begun to be used ``off-label'' to predict the structural dynamics of a single protein, as it explores different conformations \cite{monteiro2024high}. 
 
\par At the cellular scale, work by Furasawa et al. has characterized the transciptomes of \textit{E. coli} \cite{sato2020evolutionary} that have accumulated mutations during long term evolution experiments. They find that the gene expression state of these cells --- that differ through mutations --- lie in the same low dimensional subspace of gene expression as wild type cells exposed to environmental stresses and different nutrient conditions (Fig. \ref{phenocopy_fig}c)\cite{sato2020evolutionary}.
\par Recent work has extensively characterized the effect of mutations and drug perturbations on cell shape, in a high-throughput manner \cite{chandrasekaran2024three}. This work established idiosyncratic correspondences exists in the effect on cell shape, and an important future prospect is the quantification of the dimensionality of the spaces. We posit that phenocopying should be evident in this context as well. Future work should look for a common low dimensional manifold that describes the changes to cell shape from both external stressors and mutational changes. 

\par In an ecological context, as noted above, species abundances have been shown to vary along coordinated modes of covariation \cite{hekstra2012contingency,frentz2015strongly}. If phenocopying were to manifest in this context, by analogy it would mean that both environmental changes and the insertion of new species would both push the ecological system along the same low dimensional modes of coordination in
species abundance space (Fig. \ref{phenocopy_fig}d). However, experimental studies of epistasis in ecological systems have so far assumed a fixed environment; phenocopying in ecological systems thus far remains unexplored.

In the developmental context, phenocopying was the starting point for studies of \emph{assimilation} by Waddington and others \cite{waddington1953genetic, raju2023theoretical, pigliucci2010elements,waddington2014strategy}. Assimilation is a distinct hypothesis that asks whether environmentally induced phenotypes (i.e., `phenocopies') can then become genetically encoded. 
While we have argued that phenocopying is a useful and observed phenomenon on other scales in biology, it is unclear if assimilation has relevance to the examples we have discussed.  Another distinction from the developmental context - the perturbations are typically understood as large perturbations that access non-linear (e.g., as depicted in Waddington's famous landscape with multiple distinct valleys \cite{waddington2014strategy})

Finally, phenocopying-like phenomena have been explored in machine learning\cite{feng2023activity,boukacem2024waddington}, where \cite{feng2023activity} demonstrated a duality linking changes in synaptic weights in a neural network to changes in activity in that network, analagous to the mutation-environmental change correspondence.

\subsection{Dual buffering}
\begin{figure}[h]
\includegraphics[width=4in]{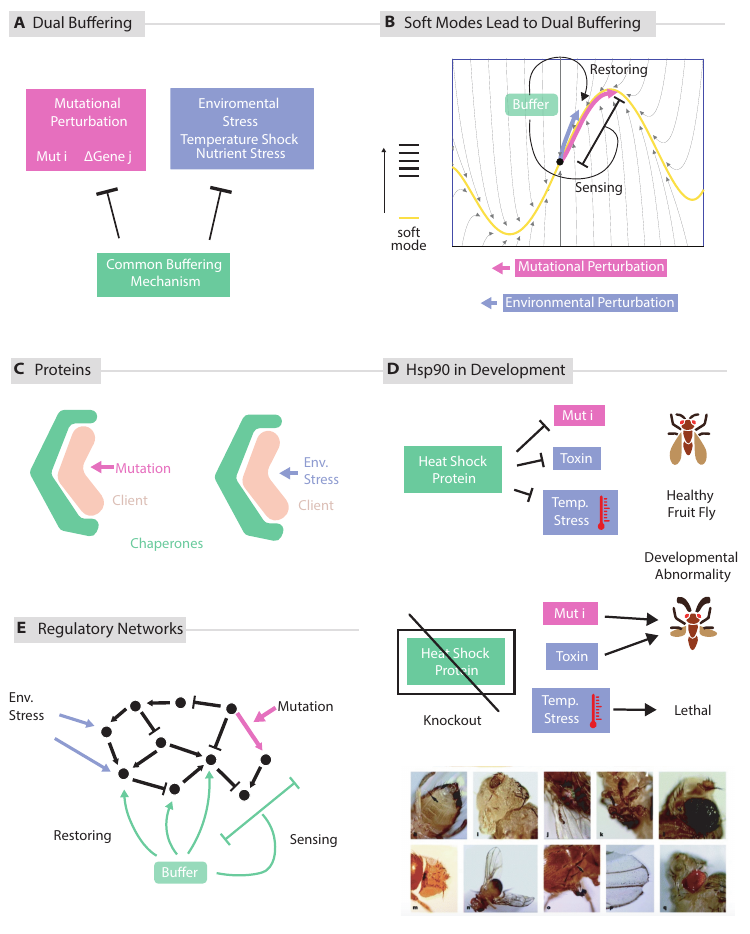}
\caption{\textbf{Soft modes predict dual buffering.} (a) Soft modes channel the effect of both mutational and environmental perturbations along a common low dimensional space. Thus a stress response mechanism that responds to state changes due to environment perturbations will be able to effectively buffer the impact of mutational perturbations as well. (b) For instance, chaperones that alleviate protein misfolding or aggregation due to environmental conditions might also be expected to fix mutation-induced misfolding. (c) Buffering in development: Hsp90, a canonical heat shock response protein, has been shown to also buffer the effects of latent, heritable genetic variation. Mutations with mild or no phenotype in the presence of Hsp90 lead to morphological variation in development in the absence of Hsp90 \cite{rutherford1998hsp90}. (d) Genes identified to mitigate the effect of knocking out any one of a large number of other genes (e.g., as seen in a double knockout library) might be expected to mitigate fitness costs of many different environments (e.g., as seen in a knockout library evaluated in multiple environments). This may be testable with, e.g., synthetic gene array technology \cite{costanzo2016global,costanzo2021environmental}. 
}
\label{buffering_fig}
\end{figure}

Soft modes imply that broad stress response mechanisms, evolved to buffer environmental perturbations, will also function as `mutational buffers' that alleviate the deleterious impact of a broad spectrum of mutations (Fig. \ref{buffering_fig}a). Such mutational buffers have been speculated to increase cryptic genetic variation in populations, increasing the odds of survival in changing environments \cite{rutherford1998hsp90,yeyati2007hsp90}.

To see why soft modes imply such a dual buffering role for stress response mechanisms, consider a stress response mechanism evolved to deal with an environmental perturbation $\delta \vec{e}$; the change in state due to $\delta \vec{e}$ is $\delta_e \vec{x} \approx \sum_i \frac{\textbf{J}_e \delta \vec{ e} \cdot \vec{v_i}}{\lambda_i} \vec{v}_i$. A stress response mechanism needs to monitor (or respond to) such a displacement $\delta_e \vec{x}$ and then take appropriate actions to restore homeostasis or otherwise restore function in the cell (Fig. \ref{buffering_fig}a). 

On the other hand, mutational perturbations will perturb the cell state to $\delta_g \vec{x}\approx \sum_i \frac{\textbf{J}_g \delta \vec{ g} \cdot \vec{v_i}}{\lambda_i} \vec{v}_i$. For systems without a mode gap,  $\delta_g \vec{x}$ is generally unrelated to $\delta_e \vec{x}$ and thus a stress response mechanism evolved to respond to $\delta_e \vec{x}$ will not be useful in responding to $\delta_g \vec{x}$. However, in the presence of a large mode gap, these displacements are generally aligned, i.e., $\arccos \delta_e \vec{x} \cdot \delta_g \vec{x} \to 0$ as can be seen from the expansions in Eqns. \ref{phenocopy1} and \ref{phenocopy2} earlier. Intuitively, both environmental and mutational changes will largely push this system along the same soft mode; thus an environmental stress response mechanism would also buffer the impact of mutations in the presence of a mode gap. 

\par \textbf{Consqeuences: } The dual role of stress response mechanisms has been studied primarily in developmental contexts.  Hsp90 is a protein canonically understood to facilitate physiological robustness in the face of temperature stress \cite{cheng1992authentic}. However, Hsp90-deficient \textit{Drosophila melanogaster}    \cite{rutherford1998hsp90} display a large number of morphological deformations due to latent mutations whose deleterious impact had been effectively silenced -- or `buffered' -- by Hsp90 (Fig. \ref{buffering_fig}c) \cite{rutherford1998hsp90}. Similar results were also found in other organisms, e.g. \textit{Arabidopsis thaliana} \cite{queitsch2002hsp90} and \textit{Danio rerio} \cite{yeyati2007hsp90}. Hsp90 was dubbed a ``evolutionary capacitor'' -- its presence allows for otherwise-deleterious mutations to be present in a population; it has been argued that this pool of latent genotypic diversity may harbor mutations that improve survival in new environmental conditions \cite{rutherford1998hsp90,yeyati2007hsp90}.


We predict that such duality in buffering holds more broadly in biology, beyond development, in any context where dimensionality reduction and soft modes are relevant. For example, experiments by Costanzo et al. \cite{costanzo2016global,costanzo2021environmental} conducted high throughput growth assays on a double knockout yeast mutant library ($G \times G$), and on a single mutant library across diverse environmental stressors ($G \times E$). For such datasets and others \cite{mace2020multi}, dual buffering suggests that genes that alleviate the deletion of broad collections of other genes in the $G \times G$ data will also be genes that alleviate the stress of many environments in the $G \times E$ data (Fig. \ref{buffering_fig}d).

Molecular chaperones are predicted to be dual buffers (Fig. \ref{buffering_fig}b). 
Chaperones reduce the misfolding or aggregation of client proteins in unfavorable environmental conditions  \cite{ellis1987proteins, jarosz2010hsp90,hartl2011molecular}; however, since the number of ways misfolding is large, chaperones are thought to recognize and act by sensing a few common features of misfolding \cite{hartl2011molecular} As a consequence, chaperones are then naturally able to reduce misfolding due to mutational perturbations as well \cite{tokuriki2009chaperonin, iyengar2022groel,hartl2011molecular} and thus reduce the deleterious impact of mutations in client proteins. Thus chaperones can also serve as a `mutational buffer' that allows for more genetic diversity in their client proteins \cite{jarosz2010hsp90}. 

The dual buffering role has yet to be investigated on ecological scales. If the presence of certain species allow ecosystems to be more robust to invasion by broad classes of invasive species\cite{mickalide2019higher,kinnunen2016conceptual,mallon2015microbial}, one could speculate that the mechanism of protection is through action on a low dimensional latent variable such as a resource level rather than, e.g., recognizing and responding to a highly specific molecular fingerprint of one invasive species\cite{mallon2015microbial}. Such protection mechanisms might then also provide robustness against systemic environmental changes (e.g., a range of temperatures or nutrient supply levels \cite{stenuit2015deciphering}). 

\subsection{Global Epistasis:}  

\begin{figure}[h]
\includegraphics[width=0.68\linewidth]{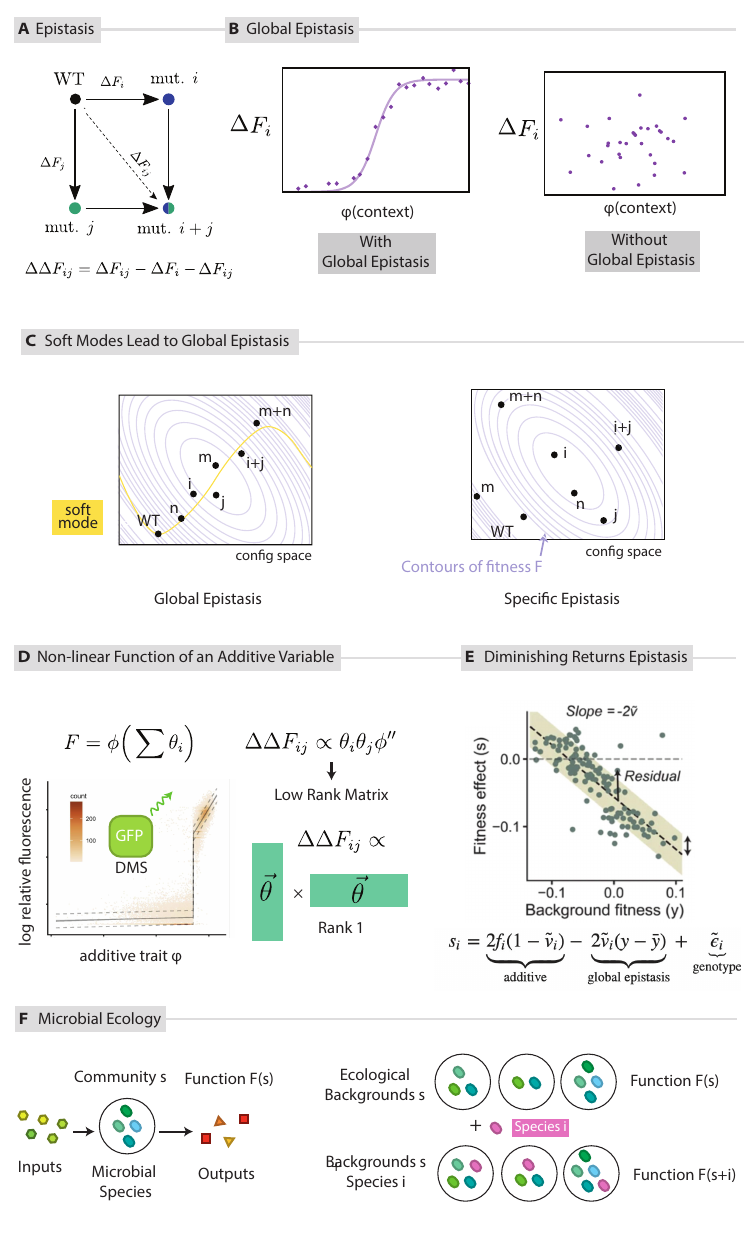}
\caption{\textbf{Soft modes predict low dimensional patterns in epistasis.} (a) Pairwise epistasis is defined  as the deviation of the sum of the independent effects of mutations \textit{i} and \textit{j} from the combined effect of both together. (b) In \textit{global} epistasis, the dependence of the effect of mutation $i$ on its genomic context can be predicted by a low dimensional variable of the system $\phi$. This dimensionality reduction can reflect in relationships between seemingly independent epistatic coefficients between different pairs of sites (e.g., low rank epistasis\cite{otwinowski2018inferring,husain2020physical}) or through unexpected statistical relationships \cite{ kryazhimskiy2014global}.
(c) Soft modes predict low rank epistasis.
Examples: 
(d) Low rank epistasis in proteins
(e) Statistical global epistasis in genomes
(f) Global epistasis has been observed in microbial ecologies, by building a low dimensional model of the functional effect $\Delta F$ of adding species $i$ to an ecological background $s$}
\label{epistasis_fig}
\end{figure}

As discussed earlier, epistasis is when the effect of a mutation is dependent on its context. In general, the fitness as a function of sequence can be expressed as a sum of epistatic contributions 
\begin{equation}\label{globalpistasiseq}
    F(\{s_i\}) = F_{WT} + \sum_i \Delta F_i s_i +  \sum _{i,j} \Delta \Delta F_{ij} s_i s_j + \sum_{i,j,k} \Delta \Delta \Delta F_{ijk} s_i s_j s_k + ...
\end{equation}
where $s_i = 0,1$ is the genotype (and $s_i= 0 $ for all $i$ represents wild type) and
where $\Delta \Delta F_{ij}$ are epistatic coefficients capturing non-linear interactions between mutations at different sites (Fig. \ref{epistasis_fig}a). Note that we are discussing epistasis with respect to a reference WT sequence; see \cite{poelwijk2016context,park2023simplicity} for a treatment of background-averaged epistasis.

In global epistasis\cite{otwinowski2018inferring,starr2016epistasis,olson2014comprehensive,kryazhimskiy2014global,reddy2021global,johnson2023epistasis,husain2020physical}, there are global patterns that dictate these many non-linear interactions. In one conception, global epistasis occurs when fitness $F$ is a non-linear function $\phi(g)$ of a global variable $g$ (Fig. \ref{epistasis_fig}b), which is itself a linear function of many different kinds of mutations: 
\begin{equation}\label{Fphi}
    F(\{s_i\}) = \phi(g(\{s_i\})) + \epsilon
\end{equation}
\begin{equation}\label{gsum}
g = \sum_{i} \theta_i s_i
\end{equation}

Soft modes are one generic, simple way to explain the origin of such epistasis (Fig. \ref{epistasis_fig}c). In a system with a large mode gap, displacement along the soft mode can act as the linear trait $g$ in Eq. \ref{Fphi} and Eq. \ref{gsum} above \cite{husain2020physical}. To see this, 
 
note, as earlier, that the change in state of the system in response to a set of mutations $\{\vec{\delta g_j}\}$ will take the form
\begin{equation}
    \delta \vec{x_j} \approx  \frac{J_g \delta \vec{g_j} \cdot \vec{v_1}}{\lambda_1} \vec{v}_1 
\end{equation}
As a consequence, the epistatic expansion of $F$ reduces to:
\begin{equation}
    F(\{s_j\}) = F(\vec{x_0} + \sum_j \delta \vec{x_j}) \approx F\Bigg(\vec{x_0} + \Big(\sum_j \frac{J_g \delta \vec{g_j} \cdot \vec{v_1}}{\lambda_1}\Big) \vec{v_1}\Bigg) = \phi\Big(\sum_j \frac{J_g \delta \vec{g_j} \cdot \vec{v_1}}{\lambda_1}\Big) = \phi \Big(\sum_j \theta_j s_j\Big)
\end{equation}
Even if $F$ is a function of the high dimensional state $\vec{x}$ of the system, in the presence of soft modes, deformations due to mutations only explore a lower dimensional space.  In this case, the epistatic coefficients in Eqn.\ref{globalpistasiseq} are no longer independent but are instead low rank matrices and tensors (Fig. \ref{epistasis_fig}d),
\begin{equation}
    \Delta F_{i}  \approx \phi'\theta_i
\end{equation}
\begin{equation}
    \Delta \Delta F_{ij}  \approx \phi''\theta_i\theta_j
\end{equation}
\begin{equation}
    \Delta \Delta \Delta F_{ijk}    \approx \phi'''\theta_i\theta_j\theta_k
\end{equation}
with $$\theta_j = \frac{J_g \delta \vec{g_j} \cdot \vec{v_1}}{\lambda_1}$$ 
If there are two soft modes, then it can be shown that 
$\Delta \Delta F_{ij} \propto c_1 \theta_i \theta_j + c_2 \mu_i \mu_j$ where $\mu$ represents displacement along the second of the two soft modes, and $\Delta \Delta F_{ij}$ is a rank 2 matrix. Thus, in the presence of soft modes, systems can show strong epistasis, but epistatic coefficients are not independent.

\textbf{Consequences:} Epistasis has the potential to make fitness landscapes highly rugged, trapping trajectories at local maxima and making them hard to predict. In contrast, when epistasis is global -- mediated by a global, low dimensional variable, the fitness landscape may be easier to navigate, and the low dimensional variable can facilitate the prediction of trajectories \cite{kryazhimskiy2014global,husain2020physical}. \cite{kryazhimskiy2014global} argued that knowledge of the state of the low dimensional variable they believe is mediating global epistasis of the founder genotype powerfully predicts its evolutionary trajectory in a directed evolution experiment. Similarly, \cite{diaz2024global}  shows how simple statistical models analogous to global epistasis can predict how adding a new species to a microbial community will affect its function (Fig. \ref{epistasis_fig}e). They argue that these models offer an avenue to predict and optimize microbial community function based on species composition.  

\section{Why soft modes?}

This review has stayed agnostic to why a soft mode might exist in an evolved biological system in the first place and focused on non-trivial consequences (phenocopying, dual buffering, global epistasis) without ascribing any fitness value to any of these. For completeness, we briefly review several reasons why soft modes might arise in the first place. 

\subsection{Direct functional benefits of soft modes}

The most parsimonious explanation for `why soft modes' is that soft modes enable a direct function. 

\textbf{Allosteric communication:} soft modes have been shown to facilitate allostery in proteins, i.e., communication between distal residues in the protein \cite{mitchell2016strain,rocks2017designing,campitelli2020role,ravasio2019mechanics,raman2016origins}. 
Analogously, a soft mode in a gene regulatory network might facilitate changes in function across the network through the regulation of one or a small number of components in the network. 

\textbf{Internal models of the external world:} Biological systems, even those without neurons, can be seen as \emph{learning} statistical structure of environments seen over their evolutionary environment\cite{gunawardena2022learning,mayer2019well}. If the effective diversity of environmental states that is \emph{relevant} to the system is relatively low dimensional, the internal state of the system that tracks or responds to the environment will match its dimensionality \cite{friedlander2015evolution,furusawa2018formation}. Stiffening  irrelevant dimensions enables the system to couple selectively to informative components of external signals while project out irrelevant noise in those signals - e.g., circadian clocks structured as limit cycle attractors allow organisms keep track of the phase of the day-night cycle while projecting out irrelevant intra-day \emph{amplitude} fluctuations in light levels\cite{pittayakanchit2018biophysical,monti2018robustness}. Mathematically similar information processing has been explored in the context of place cell attractors in neuroscience\cite{yoon2013specific}.

\textbf{Environmental robustness:} Channeling the impact of diverse environmental perturbations down a soft mode might enable a low dimensional control mechanism to respond to environmental stresses\cite{furusawa2018formation,waddington2014strategy}, as discussed in the section on dual buffering.

\subsection{Indirect evolvability benefits of soft modes}

Many of the consequences of soft modes spelled out in this review do not provide a direct fitness benefit to an individual but instead provide `evolvability' benefits to a lineage. `Evolvability'\cite{wagner2008robustness} is often defined as the ability of a lineage to undergo evolution more effectively, rather than a simple survival benefit to an individual. When seeking to establish an evolvability-based selection argument for soft modes, it is important to remove direct functional benefits as a confounding factor; e.g., see \cite{reynolds2011hot,raman2016origins} on the evolution of allostery in a context without a need for allosteric regulation. 
\textbf{Channeled variation:} One apparent evolvability benefit of soft modes is the channeling of variation that natural selection can act on. The outcome of evolution can be dramatically altered by changing the kind of variation available \cite{good2015impact,card2019historical,lande1983measurement}. If the eigenvector corresponding to soft modes is aligned to the direction of phenotypic change demanded by a changing environment, a system with soft modes will more quickly generate beneficial mutations upon a change in environment \cite{furusawa2018formation}.
\par

\textbf{Reduced epistasis:} soft modes could also potentially reduce the ruggedness of fitness landscapes. For example, fitness landscapes described by global epistasis can be much less rugged, as discussed earlier. Rank-1 epistasis matrices correspond to a smooth Hopfield-like landscape, while full rank matrices can result in glassy landscapes \cite{husain2020physical}. `Protein sectors', a theory of co-evolving units, also predicts soft modes in proteins due to a pressure to change ligand specificity over time \cite{raman2016origins, halabi2009protein}. 
\par
\textbf{Mutational robustness:} A related evolvability benefit of soft modes is mutational robustness as discussed in the dual buffering section. Mutational buffers or `capacitors' can exploit soft modes to maintain larger standing genetic variation in a population which might be beneficial in changing environments \cite{masel2006cryptic}. 

\section{Other Perspectives}
Here, we briefly discuss other perspectives on low dimensionality in biological systems outside of the soft modes framework.

\subsection{Incidental Low Dimensionality and an Alternative Model of Global Epistasis}
Biological systems might appear low dimensional for entirely mathematical, statistical or data limitation reasons. 

For example, global epistasis patterns seen in data \cite{reddy2021global,kryazhimskiy2014global,petti2023inferring,ardell2024environment} has been argued to potentially arise entirely for statistical reasons. Models in which interaction coefficients are drawn from specific random ensembles and the number of interacting degrees of freedom is large predict global epistasis patterns seen in data {(in a phenomena sometimes referred to as "diminishing returns epistasis")}. \cite{reddy2021global,petti2023inferring}. We note that the precise definition of global epistasis differs between the works concerned with the whole genome and works concerned with single or a few proteins. More critically, the generic statistical explanation is distinct from the soft mode explanation discussed earlier since the latter relies on a special property - a separation of timescales - that might be found in evolved systems but not random ones. 

It has also been shown that dimensionality reduction of biological data is possible, despite the absence of any underlying low dimensional constraints on the system, due to statistical properties of high dimensional systems \cite{eckmann2021dimensional,moore2018high}. 
For example, a Gaussian random walk in a high dimension space will appear low dimensional when subject to dimensionality reduction methodologies such as PCA or Multi-Dimensional Scaling \cite{moore2018high}.

\subsection{Low Dimensionality and Machine Learning}

In recent years, machine learning methods have proved very effective at solving certain kinds of problems involving high dimensional biological data \cite{greener2022guide,zitnik2019machine,jumper2021highly}. Some have argued that the success of these methodologies relies on the effective low dimensionality of data \cite{heimberg2016low}. 

In some cases, soft modes may explain the success of these methodologies. AlphaFold, which predicts protein structure from sequence, relies on evolutionary data: to predict the structure of a sequence, it relies on an aligned set of structures and sequences for homologous proteins \cite{jumper2021highly}. As discussed above, soft modes constrain the dimensionality of the structural variation of proteins, and thus likely enable AlphaFold's success.

\subsection{Modularity}
Modularity is often cited as a cause of low dimensionality in biological systems \cite{alon2019introduction}, in particular, in the context of gene regulatory networks\cite{heimberg2016low,segal2003module,bergmann2003iterative}. Indeed, modular organization is an ubiquitous feature of biological systems -- from gene regulatory network architecture to the very fact of multicellularity and to the existence of organs in larger multicellular organisms \cite{kashtan2005spontaneous}. A modular network architecture of a gene regulatory network straightforwardly facilitates low dimensionality -- if a single upstream node controls many downstream ones, naturally, the system will vary in a low dimensional manner as the upstream node changes. 
While modularity and the soft mode framework intersect in many ways, there are differences. Fundamentally, modularity is typically formulated as a statement about the topology of a network, while soft modes are about the dynamics and thus agnostic to topology. Modular topologies of networks might often result in soft modes but the converse is not true. 
Consequently, the three central consequences of soft modes spelled out in this review do not require modularity but might be enabled by it.

\section{Conclusion}
In this review, we began by presenting a salient set of examples of low dimensionality across scales in biological systems. We showed how slow dynamical modes are a powerful explanatory framework, beyond the context of developmental biology where such ideas were first introduced into the life sciences by Waddington. We identified three key predictions about the behavior of biological systems where low dimensionality is caused by soft modes  -- phenocopying, dual buffering, and global epistasis -- and showed how they manifest in empirical work across scales. We concluded with a discussion of the possible origins of soft modes in biological systems as well as alternate conceptions of low dimensionality that have shaped scientific discourse. Understanding the nature and origin of low dimensionality in biological systems helps us tackle decades-old problems in the life sciences, like the predictability of evolution and the genotype-phenotype map, and is of critical importance given that low dimensionality underlies the effectiveness of modern tools in computational biology. Soft modes are not a universal framework, but a powerful one that is likely at play in many systems across scales.


\section*{DISCLOSURE STATEMENT}
The authors are not aware of any affiliations, memberships, funding, or financial holdings that
might be perceived as affecting the objectivity of this review. 

\section*{ACKNOWLEDGMENTS}
We are grateful to Vedant Sachdeva, Rama Ranganathan, Jane Kondev, Rob Phillips, Abigail Skwara, Pankaj Mehta,  Seppe Kuehn, Mikhail Tikhonov, Terry Hwa, Lauren McGough, the Murugan lab, the Chan-Zuckerberg theory group for helpful discussions. This work was supported by the Chan-Zuckerberg Initiative, National Science Foundation through the Center for Living Systems (grant no.  2317138) and the NIGMS of the NIH under award number R35GM151211. 

%


\bibliography{bib}

\end{document}